\documentclass[floatfix]{revtex4}
\usepackage{epsfig,bm,latexsym,amssymb,amsfonts,amsmath,graphicx}
\usepackage{amsmath}
\usepackage{amsfonts}
\usepackage{epsf}
\def\beq{\begin{equation}}
\def\eeq{\end{equation}}

\newcommand{\mbf}{\mathbf}

\usepackage[usenames]{color}
\definecolor{Red}{rgb}{1,0,0}

\begin{document}
\title{Optimal experimental design in an EGFR signaling and 
down-regulation model}

\author{Fergal P. Casey$^1$, Dan Baird$^2$, Qiyu Feng$^3$, Ryan N. Gutenkunst$^4$, Joshua J. Waterfall$^4$, Christopher R. Myers$^5$, Kevin S. Brown$^6$, Richard A. Cerione$^3$, James P. Sethna$^4$} 

\affiliation{$^1$ Center for Applied Mathematics, Cornell University, Ithaca, NY 14853, USA}
\affiliation{$^2$Department of Cellular and Molecular Medicine and the Howard Hughes Medical Institute, University of California at San Diego, La Jolla, CA 92093, USA} 
\affiliation{$^3$Department of Molecular Medicine, College of Veterinary Medicine, Cornell University, NY 14853, USA} 
\affiliation{$^4$ Laboratory of Atomic and Solid State Physics, Cornell University, Ithaca, NY 14853, USA}
\affiliation{$^5$Cornell Theory Center, Cornell University, Ithaca, NY 14853, USA} 
\affiliation{$^6$Department of Molecular and Cellular Biology, Harvard University, Cambridge, MA 02138, USA} 

\begin{abstract}
We apply the methods of optimal experimental design to a differential equation model for 
epidermal growth factor receptor (EGFR) signaling, trafficking, and down-regulation. 
The model incorporates the role of a recently discovered protein 
complex made up of the E3 ubiquitin ligase, Cbl, the guanine exchange 
factor (GEF), Cool-1 ($\beta$-Pix), and the Rho family G protein Cdc42. The complex has been 
suggested to be important in disrupting receptor down-regulation~\cite{3T3expt3,qiyulatest}.
We demonstrate that the model interactions can accurately reproduce the experimental 
observations, that they can be used to make predictions with accompanying uncertainties, 
and that we can apply ideas of optimal experimental design to suggest new experiments 
that reduce the uncertainty on unmeasurable components of the system.
\end{abstract}

\maketitle

\section{Introduction}
The epidermal growth factor receptor (EGFR) is a transmembrane tyrosine kinase receptor
which becomes activated upon binding of its ligand, epidermal growth factor (EGF), 
and signals via phosphorylation of various effectors~\cite{Carpenter}. 
Besides sending signals to downstream effectors, the activated EGFR also will initialize
endocytosis which is followed by either degradation or recycling of the receptor.
These are the normal receptor down-regulation processes.
Persistence of activated receptor on the cell surface can lead to aberrant signaling and 
transformation of cells~\cite{wells}. 
In addition, a variety of tumor cells exhibit overexpressed or hyperactivated EGF 
receptor~\cite{lungcancer, breastcancer}, indicative of the failure of normal receptor 
down-regulation. 

\begin{figure}[hbtp]
\begin{center}
\epsfxsize=7.0in
\epsfbox{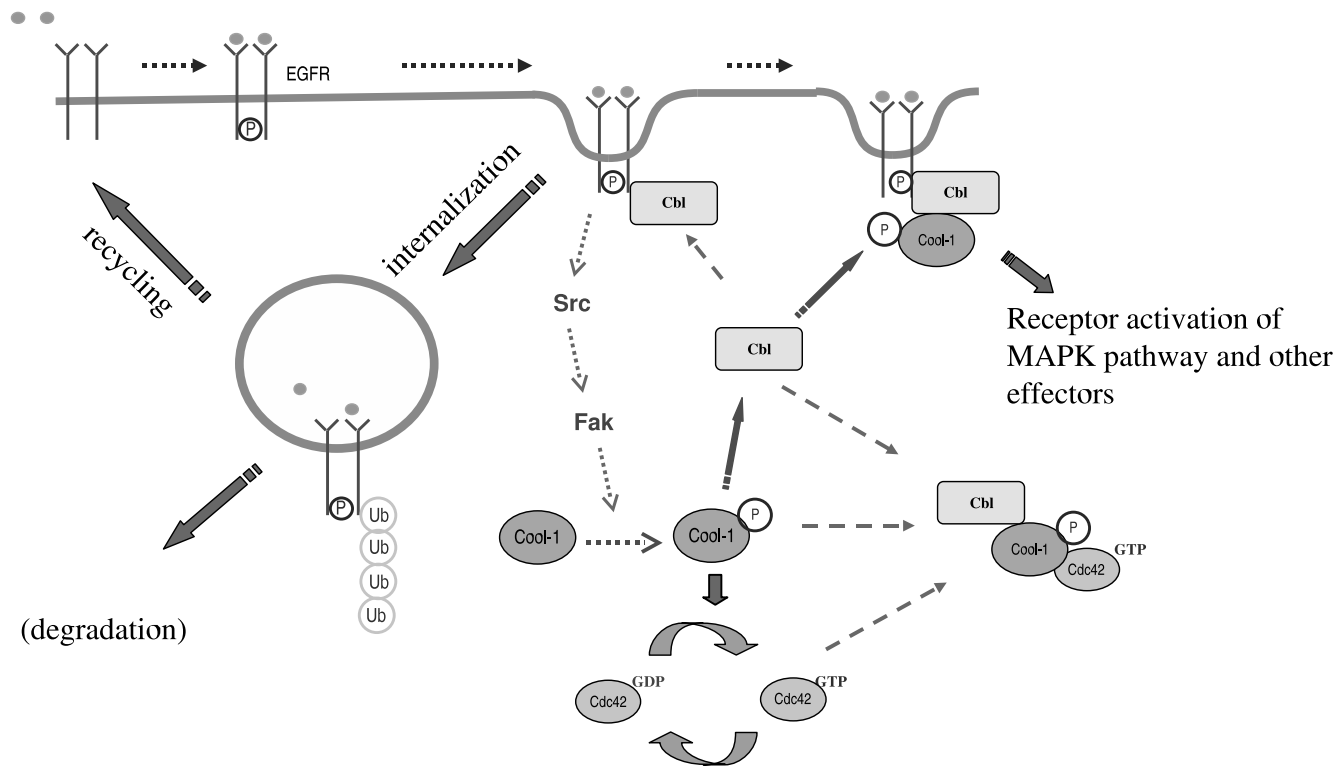}
\caption{Schematic diagram showing the set of interactions in the model of EGFR 
signaling, endocytosis and down-regulation (see also~\cite{qiyulatest}). Phosphatases
are not shown.}
\label{fig:EGFRnetwork}
\end{center}
\end{figure}

We concern ourselves with building a mathematical model of the receptor
endocytosis, recycling, degradation and signaling processes that can reproduce 
experimental data and incorporates the effects of regulating proteins
that themselves become active after EGF stimulation. 
The schematic for the model is shown in 
Fig.~\ref{fig:EGFRnetwork}.
In particular, we examine the roles of the GEF, Cool-1, and the GTPase, Cdc42 that have 
recently been discovered to be important for EGFR homeostasis~\cite{qiyulatest,3T3expt3}
through their interaction with the E3 ubiquitin ligase, Cbl. There is evidence for 
two interaction mechanisms which disrupt the normal receptor down-regulation. 

The first mechanism involves the formation of a complex between active Cool-1, active
Cdc42 and Cbl. After activation of the receptor, Cool-1 becomes phosphorylated 
through a Src - FAK phosphorylation cascade. Phosphorylated Cool-1 has GEF activity 
and in turn activates Cdc42 by catalyzing the exchange of GDP for GTP. Unlike other GEFs however,
activated Cool-1 can remain bound to its target, Cdc42,~\cite{3T3expt3} 
and can then form a complex with Cbl (mediated through Cool-1 binding), effectively 
sequestering Cbl from the receptor.
Therefore the internalization and degradation of the receptor is inhibited and its growth signal
is maintained. (We use the ERK pathway as a readout on the receptor mitogenic signal.) 
The second mechanism is based on the findings of~\cite{qiyulatest} that activated 
Cool-1 can directly bind to Cbl on the receptor and block endocytosis in a manner we 
hypothesize be analogous to the action of Sprouty2~\cite{sprouty}.  

To maintain normal receptor signaling, we postulate it is crucial that deactivation of Cool-1
and subsequent dissociation of the Cbl, Cool-1 and Cdc42 complex occur. Then Cbl can 
induce receptor internalization and ubiquitin tag it for degradation in the lysosome. 
Internalized receptor lacking ubiquitin moieties can be returned to the cell surface from 
the early endosome via the recycling pathway. 

The role of Cbl in the degradation mechanism for the receptor
has been understood for some time~\cite{cblrole,cblrole2,y1045importantForDeg}. However, its 
function in mediating endocytosis still remains controversial (e.g.~\cite{grb2importantForEndo,
SorkinCblneeded,YardenMonoUbi,cblnotneededforEndo,difiore}) as the receptor can be internalized
through more than one endocytic pathway. We do not address that issue here but 
rather we assume in our model that Cbl association and activation is necessary for 
endocytosis, whether through a CIN85-endophilin interaction~\cite{dikic} or through 
ubiquitination of the receptor~\cite{difiore} and therefore we do not include a separate 
Cbl-independent endocytosis pathway. 
The overall set of these protein-protein interactions is summarized in 
Fig.~\ref{fig:EGFRnetwork} (we also incorporate phosphatases in the model to 
act on the various phosphorylated species, but this is not shown in the network figure).
There is a significant overlap between 
our model and previous models of EGF receptor signaling and/or trafficking, 
~\cite{schoeberl,kholodenko,integratedWileyModel,blinov}. Since we wish to focus
on the role of the Cool-1/Cdc42 proteins within the network and to demonstrate the 
utility of optimal experimental design, we leave out some of the known intermediate  
reactions involved in the MAPK and EGFR-Src activation pathways, preferring 
a ``lumped'' description which is more computationally manageable.

The goals of this manuscript are to demonstrate how a modeling approach can be
used to  
\begin{trivlist}
\item[{(a)}] refine the necessary set of interactions in the biological network, 
\item[{(b)}] make predictions on unmeasured components of the system with good precision and 
\item[{(c)}] reduce the prediction uncertainty on components that are difficult to measure
directly, by using the methods of optimal experimental design.
\end{trivlist}

\section{Methods}
\subsection{Mathematical model, parameter and prediction uncertainties}
Before we introduce the algorithms needed to address the design question, we define
the model and data in more detail. Our differential equation model for EGFR signaling 
and down-regulation contains 56 unknown biochemical constants: 53 unknown rate and Michaelis Menten 
constants (where they can be found, initial estimates were drawn from the literature), 
and 3 unknown initial conditions which we found useful to vary. 
The dynamical variables are comprised of 41 separate chemical species, including complexes. 
The data consist entirely of time series in the form of Western blots. (The
data both come from the lab of the co-authors and from the literature, see supplementary
information for details.)  
We have been careful to select data only on NIH-3T3 cells, and in 
experimental conditions where the cell has been serum-starved prior to EGF stimulation, to
prevent activation events not related to the EGFR ligand binding. Most of the time series
data are over a period less than a few hours which allows us to ignore transcriptional processes.

Since we have no information on most of the biochemical constants, we must infer them from
the data. Therefore we optimize a cost function which measures the discrepancy
of simulated data from the real data, 
\begin{equation}
C(\theta) = \sum_{\alpha=1}^{D} 
\sum_{i=1}^{m_{\alpha}} 
\left( \frac{y_{\alpha}(t_{\alpha i},\theta)-d_{\alpha i}}{\sigma_{\alpha i}} \right)^2
\label{cost}
\end{equation}
where $\alpha$ is an index on the $D$ measured species, $m_{\alpha}$ is the number of time points on species 
$\alpha$, $y$ is the trajectory of the differential equation model, $\theta$ is a vector of the 
logarithm of the biochemical constants, 
$d_{\alpha i}$ is the measured value at time $t_{\alpha i}$ for 
species $\alpha$ and $\sigma_{\alpha i}$ is the error on the measured value.
In other words, we have a standard weighted least squares problem to reduce
the discrepancy of the model output to the data by varying $\theta$. 
(We use the logarithm of the biochemical constants as it allows us to apply an 
unconstrained optimization method while maintaining the positivity constraint 
and it removes the discrepancies between biochemical values that have naturally different scales
in the problem). As absolute numbers of proteins in the network cannot be accurately 
measured, data sets measuring activities of proteins are fit up to an arbitrary 
multiplicative scale factor, which adds parameters to the model not
of direct inferential interest (\emph{nuisance parameters}).
Where the relative quantity of a species can be measured (normalized by the level before 
EGF stimulation for example), 
the output of the differential equations are similarly scaled by an appropriate 
common factor. 

After the model has been successfully fit to the experimental data, we have a parameter
estimate $\hat{\theta}$ which in general will have large covariances, approximated by 
the inverse of the \emph{Fisher information matrix} (FIM). The FIM is defined as 
\begin{eqnarray}
 M & = & \mbf{E}[\partial^2 C / \partial \theta^2] \\ 
   & = & \sum_{\alpha=1}^{D} \sum_{i=1}^{m_{\alpha}} 
   \frac{1}{\sigma_{\alpha i}}
   \frac{\partial y_{\alpha}(t_{\alpha i},\theta)}{\partial \theta}^{t} \big |_{\hat{\theta}} 
   \frac{1}{\sigma_{\alpha i}}
   \frac{\partial y_{\alpha}(t_{\alpha i},\theta)}{\partial \theta} \big |_{\hat{\theta}} \\
   & = & J^{t} J \quad .  
\label{fisherinf}
\end{eqnarray}
where the expectation is over the 
distribution of errors in the data, which are assumed to be Gaussian.
The expression for the FIM above is exact when the model fits perfectly, 
{i.e.} at the best fit, the expectation of the 
residuals is zero, $\mbf{E}[y_{\alpha}(t_{\alpha i}) - d_i] = 0$. The $i^{th}$ 
parameter uncertainty is given by the square root of the $i^{th}$ diagonal element 
of the inverse FIM.
$J = \frac{1}{\sigma_{\alpha i}}
\partial y_{\alpha}(t_{\alpha i},\theta)/\partial \theta |_{\hat{\theta}} $ is the 
the sensitivity matrix of residuals with respect to parameters at the best fit and is the analog to 
the \emph{design matrix} in a linear regression setting. 
The \emph{design space} is the 
range of species $\alpha$ and of time points $t_{\alpha i}$ for which measurements could be 
taken. ($\alpha i$ is the row index of $J$.)

We can also make predictions on components of the trajectory (measured or unmeasured), 
$\hat{y}_{\beta}(t) = y_{\beta}(t,\hat{\theta})$. 
The variances on these quantities are given by 
\begin{equation}
\mbox{Var}(\hat{y}_{\beta}(t)) \approx 
\frac{\partial y_{\beta}(t,\theta)}{\partial \theta}^t|_{\hat{\theta}} M^{-1} 
\frac{\partial y_{\beta}(t,\theta)}{\partial \theta}|_{\hat{\theta}} 
\label{predvariance}
\end{equation}
The form of Eqn. \ref{predvariance} can be thought of as a combination of the underlying
parameter uncertainty, quantified by $M^{-1}$, and the linear response of the
system to the parameter uncertainty, quantified by the sensitivities.
Note that $M$ is also computed using the sensitivities of the trajectory of the differential 
equations, which we obtain by implementing the forward sensitivity equations~\cite{sundials}. 
In practice, $M$ is close to singular if we do not include some prior information on 
parameter ranges. Therefore we assume a Gaussian prior on the parameters 
centered on the best fit values and with a standard deviation of $\log(1000)$. 
(This corresponds to an approximately 1000-fold increase or decrease in the non-logarithmic 
best fit biochemical values.)

We recognize there can be other sources of uncertainty in predictions, 
for example if the dynamics of the system are modeled stochastically or if there is 
model uncertainty that needs to be taken into account. The former is not relevant 
here as the measurements we fit are not on the single cell level, but rather the 
average of large populations of cells. The latter is certainly of interest but we
choose an approach where model errors are corrected during the fitting and validation
process, rather than included \emph{a priori} in the model definition.

Given the approximate nature of variance estimates derived from the Fisher information matrix
and the linearized model response, we supplemented these calculations with a computationally
intensive Bayesian Markov Chain Monte Carlo (MCMC) method to compute credible intervals 
for the predictions we make on the model (see supplementary material). The estimates from the 
Bayesian MCMC approach are in sufficient agreement with the linearized error anaylsis results 
that we believe the optimal experimental design algorithms 
introduced below are justifiably aimed at reducing the approximate uncertainties 
of Eqn.~\ref{predvariance}. Using MCMC for error estimates within the framework 
of the optimal design algorithms would be computationally infeasible.

\subsection{Optimal Experimental Design}
Optimal experimental design is a technique for deciding what
data should be collected from a given experimental system such
that quantities we wish to infer from the data can be
done so with maximum precision. Typically the network as shown in Fig.~\ref{fig:EGFRnetwork}
has components that can be measured (e.g. \emph{total} levels of 
active Cdc42, \emph{total} levels of surface receptor etc.) and components that are 
not directly measurable (e.g. levels of the triple complex comprising Cool-1, Cdc42 and Cbl). 
Therefore we can pose the question
of how to minimize the average prediction uncertainty on some unmeasurable component of 
interest by collecting data on measurable components of the system (we will use the term 
\emph{unmeasurable} loosely for the remainder to describe species that are 
between difficult and impossible to measure by standard methods).
This is just one possible
\emph{design criterion}, called V-optimality in the literature; 
other criteria involve minimizing the total parameter uncertainty in the system 
(D-optimality), minimizing the uncertainty in the least constrained direction in parameter
space (E-optimality) or minimizing the maximum uncertainty in a prediction 
(G-optimality)~\cite{Atkinson}.
Other authors~\cite{Klingmuller, Banga, optimalsamplingtimes} have focused on reducing
parameter uncertainty but we believe that complex biological models, even with 
large amounts of precise time series data, have intrinsically large parameter 
uncertainty~\cite{kevin1, Ryan,Josh}.
On the other hand, even with no extra data collection, the uncertainty on unmeasured time 
trajectories in these biological systems can be surprisingly small despite the large parameter 
uncertainty~\cite{Ryan}.

By altering the form of the matrix $J$ in Eqn.~\ref{fisherinf}, 
by measuring different species at different times, we have the possibility of 
reducing the average variance of $\hat{y}_{\beta}$, which is an integral over 
time of the quantity defined in Eqn.~\ref{predvariance}. We discuss the types of 
design and algorithms that can be used to achieve this.

A distinction must be made between \emph{starting designs} and
\emph{sequential designs}. A starting design is one in which no data has
been collected and the experimenter would like to know what design is
best to minimize a given criterion function. Within this category are two
subcategories: \emph{exact designs} and \emph{continuous designs}. 
Exact designs refer to the optimal placement of a finite number of design 
points. As the the design points need to be assigned amongst all the measurable 
species in the system the optimization problem is of a combinatorial nature.
There have been specific algorithms developed for this situation~\cite{Atkinson} 
which involve choosing some initial design with the required 
number of points and then randomly modifying it 
by doing exchanges, additions and deletions. More general 
global optimization algorithms have been applied to the problem of finding exact 
designs in differential equation and regression models~\cite{JonesWang,globOptDesign}.

Continuous designs refer to the selection of
a \emph{design measure}, $\eta$, which is equivalent to a probability density over
the design space. The advantage of assuming a continuous design is that
the criterion function can then be differentiated with respect to the
design measure and tests for optimality can be derived. 
Asymptotically, for a large number of design points the continuous and 
exact designs should coincide. For a \emph{linear} model described by 
$y = f(t)^{t} \theta + \epsilon$ where $f(t) \in \mbf{R}^{N}$ and $\epsilon$
is an error term, the FIM is 
\[ M(\eta) = \int_{\tau} f(t) f(t)^{t} \eta(x) \, dt \]
by definition of the design measure, $\eta$. However, $M$ is a symmetric $N \times
N$ matrix made up of a convex combination of the rank one symmetric matrices,
$f(t) f(t)^t$. Therefore it can be represented by a convex combination of at 
most $N (N+1)/2$ design points (from Caratheodory's Theorem) 
$x_1,\ldots,x_{N(N+1)/2}$, 
i.e. as a convex combination of delta function probability
measures on those points. In other words even continuous optimal designs for
linear models have only a finite number of design support points~\cite{Silvey}. 
In one of the approaches that follows, we will attempt to find a continuous design by
approximating the design measure by a number of finely spaced measurement points with
weights associated with each one, and we will see that a near optimal design is
in fact only supported on a small subset of those points. 

Sequential designs are more relevant to the situation we consider here: 
experimental data have already been collected and the model has already been fit. 
Therefore we can get an initial estimate for the parameters in the 
system and we can evaluate the FIM. Suppose that
the current design already has $n$ points and the current FIM is 
$M_n = J_n^{t} J_n$.
The effect of adding the $(n+1)^{th}$ design point 
({e.g.} $y_{\alpha}$ at time point $t_{\alpha i}$) merely adds a single row to $J_n$. 
Therefore the new FIM is the old FIM plus a rank one update:
\begin{equation*}
M_{n+1} = J_{n+1}^{t} J_{n+1} = J_n^{t} J_n + 
	\frac{\partial y_{\alpha}(t_{\alpha i})}{\partial \theta} \big |_{\hat{\theta}}
	\frac{\partial y_{\alpha}(t_{\alpha i})}{\partial \theta}^{t} \big |_{\hat{\theta}} \quad .
\end{equation*}
The new inverse FIM is also a sum of terms 
(by applying the Sherman-Woodbury-Morrison formula~\cite{VanLoan}): one involving 
the inverse of the old FIM and the other involving the sensitivity vector at the new point, 
$\partial y_{\alpha}(t_{\alpha i}) / \partial \theta |_{\hat{\theta}}$, so evaluating 
Eqn. \ref{predvariance} for a large number of proposed measurements is computationally
inexpensive. 

We take an approach which is a combination of continuous design and sequential 
design: assume that some initial experiments have already been carried out and 
we have an FIM for the system. We will then define a cost function 
$K(\alpha,t_{\alpha i})$ based on the integral of Eqn. \ref{predvariance}
and minimize it with respect to $\alpha$ and $t_{\alpha i}$. Initially the minimization
looks for the best single data point to reduce the uncertainty (a sequential 
design method). Once we know for which species the data needs to be collected, we 
can then place many potential measurements on that species with associated weights and 
minimize over the weights (to mimic continuous design methods where the set of weights
is the approximate design measure).

\section{Results}
\subsection{Model refinements}
\begin{figure}[hbtp]
\begin{center}
\epsfxsize=3.0in
\epsfbox{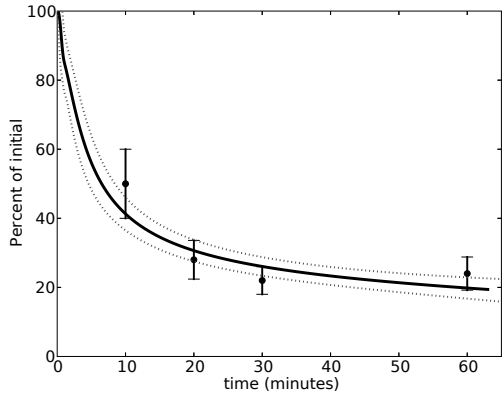}
\caption{Example of experimental fit and uncertainties around the best fit
trajectory (dotted lines) for total surface receptor (in experiment 1 in 
supplementary information).}
\label{fig:samplepred}
\end{center}
\end{figure}

The model was fit to 11 data sets, all Western blot data that describe various
signaling, internalization and degradation events that are triggered after  
receptor activation by ligand, see supplementary information for the full set of 
fitted time series and description of experiments.
As an example of a experimental fit with uncertainties, we show in 
Fig.~\ref{fig:samplepred} the best fit time course and 
standard deviation for total surface receptor from one of the experiments for which 
data was included in the model (experiment 1 in supplementary information).

During the iterative process of fitting and model refinement we discovered 
certain interactions and model parameters had to be adjusted to be consistent with
the experimentally observed behavior. We briefly summarize these 
adjustments below.

(1) It appears necessary to incorporate an interaction to allow the triple complex to be 
dissociated by a dephosphorylation reaction. In particular a reaction was 
needed whereby Cool-1 within the complex could be inactivated by its own 
dedicated phosphatase (a possible candidate already present in the system 
is SHP-2, which has been shown to dephosphorylate the related 
Sprouty protein~\cite{shp}).
Without this effect, we would not observe the complete deactivation of Cool-1 as 
it would be ``protected'' within the triple complex. Additionally, a sensitivity 
analysis to determine dominant reactions in the model identified phosphatase
reactions as important (see supplementary material).

(2) Interestingly, there is an important balance between the level of receptor and 
Cool-1 in the system to maintain the correct dynamics: if the level of receptor greatly 
exceeds the Cool-1 level, then the activated receptor will lead indirectly to 
phosphorylation of Cool-1 which in turn sustains the level of signaling receptor before
significant amounts can be endocytosed. 
It is also essential that the protein level of Cdc42 in the system be sufficiently 
high, approximately balanced with the Cbl levels as they both come together in 
the triple complex.
If this was not so, the greatly reduced Erk pathway signaling we see in 
the data set for the Cdc42 knockdown would not be possible to reproduce (supplementary
information, experiment 8). Of course, Cdc42 is involved in many other cellular 
processes, so what is actually important here is the amount available to participate in 
interactions with Cbl.

(3) The F28L fast cycling (hyperactive) mutant of Cdc42 has the ability to 
delay endogenous receptor down-regulation for many hours beyond wild type cells (see
experiment 5 in the supplementary information). 
This is only possible if the binding affinity of active Cdc42 to the Cool-1-Cbl 
complex is strong enough to deplete the levels of the latter and force the forward 
binding reaction of Cbl to activated Cool-1. This provides a mechanism
to sequester more of the Cbl protein (in both the triple complex and the Cool-1-Cbl
complex) than would otherwise be possible.

In addition to the above adjustments, we made the following observations relating to
the network dynamics and structure.

We find that given these experimental data sets, an endocytosis mechanism which is 
Cbl dependent and solely acts on activated receptors is completely consistent, although
we acknowledge that there is much controversy in the literature as to the dominant
endocytosis mechanisms and required regulators. Incorporating a Cbl independent and 
Cbl dependent mechanism in the same model does not improve the fit.
However, given only a Cbl independent endocytosis mechanism,
the model would be unable to account for the apparent saturation of the internalization
rates for overexpressed receptors (experiments 1-3 in supplementary information) 
compared to endogenous receptors (experiment 5 in supplementary information).
Therefore having a Cbl dependent pathway is convenient in explaining those experimental
observations, although any number of proteins, not in the model, could cause 
saturation in the endocytic pathway.

Despite the apparently earlier activation of Cdc42 than its putative GEF, active Cool-1, 
(see experiments 10 and 11 in supplementary information) the data still supports a 
mechanism whereby Cdc42 activation only occurs through Cool-1. The explanation of this 
effect is that the level of Cool-1 is significantly higher than 
Cdc42. Then, while only a fraction of Cool-1 is 
being activated at early times, it is still sufficient to induce substantial activation
of Cdc42. This is an example of an apparently contradictory experimental result which
only after quantitative modeling is shown still to be consistent with the proposed 
mechanism. In particular, we found there was no need to invoke another
parallel activation mechanism for Cdc42 (through Vav for example) as initially might
have been assumed.

It is in the preceding way that the interactions in the system are tuned and
modified to describe the observations, and where we get insight into the 
mechanisms of regulation. As new experimental data are collected, the model 
may have to be modified again. To decide this, the new data must be 
matched against predictions from the current model. If the predictions with uncertainties
suggest behavior contrary to the observed behavior, we would need to redefine the model. 

\subsection{Predictions}
Once we have a model which reproduces the experimental observations,  
we would like to make predictions on unmeasured or unmeasurable components 
of the system. The motivation is twofold. 
Firstly, if we make a prediction on a 
currently unmeasured component of the system which is subsequently measured, 
we have an opportunity to test the validity of our model. Secondly, if we are 
confident in the model, we may want to test a hypothesis about the role of 
an unmeasurable component in the system. If that unmeasurable component has large
uncertainties, we then need to apply the methods of experimental design to 
improve the situation. We will discuss these issues in what follows.
\subsubsection{Model validation}
\begin{figure}[hbtp]
\begin{center}
\epsfxsize=4.0in
\epsfbox{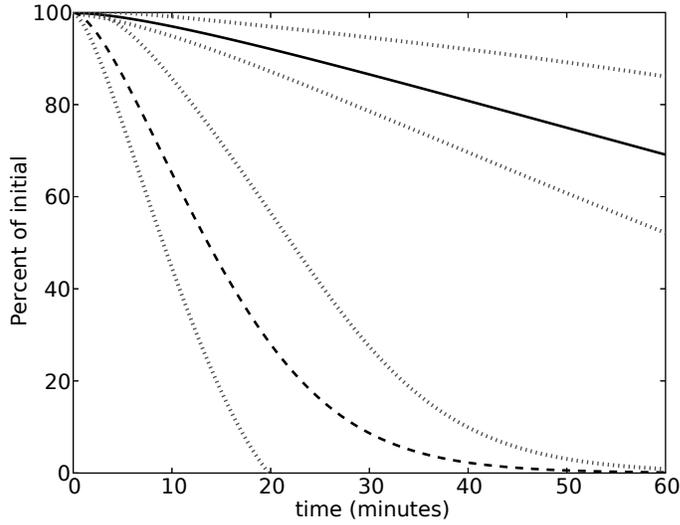}
\caption{Total surface receptor numbers after EGF stimulation in stably expressing v-Src
cells. Endogenous levels of Cool-1 (dashed curve) or overexpressed Cool-1 (solid curve).
The dotted lines show the uncertainties in each of the best fit predictions}
\label{fig:vSrcComparison}
\end{center}
\end{figure}

To first give an example of model validation, consider the qualitative 
observation in~\cite{qiyulatest} that in stably expressing v-Src cells, in conditions where
Cool-1 is overexpressed, ligand-induced receptor internalization is blocked 
compared to an endogenous Cool-1 control, for at least 60 minutes. The model is 
adjusted to simulate the conditions of these v-Src cells by making all Src in 
its active form, switching off Src inactivation and increasing the initial 
amounts 10-fold to mimic the stable transfection. We then predict the total surface 
receptor number under the two conditions and assign uncertainties using 
Eqn.~\ref{predvariance}. The results are shown in Fig.~\ref{fig:vSrcComparison}.
The qualitative observation of strong inhibition of internalization under conditions
of overexpressed Cool-1 is verified by the model. Note that in this case the 
uncertainties are small enough that we can confidently predict a large difference in
the fraction of receptors on the cell surface after 60 minutes under the two conditions.
Interestingly, the model also predicts this inhibition is much weaker in cells that 
are not stably expressing v-Src, essentially because the Cool-1 is not ``pre-activated''
and endocytosis of significant numbers of receptors can occur before the pool of Cool-1
can become phosphorylated. 
\subsubsection{Optimal design for the triple complex}
\begin{figure}[hbtp]
\begin{center}
\epsfxsize=3.0in
\epsfbox{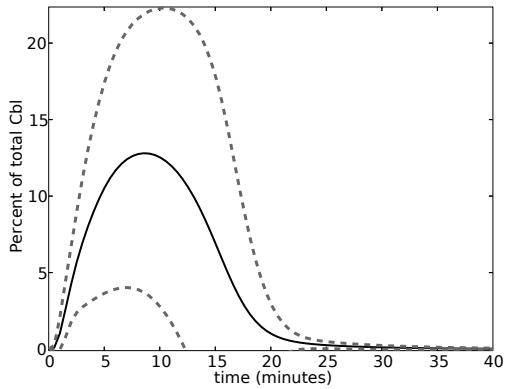}
\caption{Predictions with uncertainty on the time course of the triple complex 
consisting of active Cool-1, Cbl and active Cdc42. The quantity plotted is the 
percentage of total Cbl that is bound in the triple complex.}
\label{fig:predictions}
\end{center}
\end{figure}

Another question of interest is whether the triple complex, which appears to be
responsible for sequestering Cbl and blocks receptor down-regulation when 
Cdc42(F28L) is expressed, also forms in appreciable amounts in wild type cells. 
We would assume the answer is affirmative, as we observe a reduced downstream 
mitogenic signal from the receptor under conditions of knockdown of Cool-1 or Cdc42. 
Since the triple complex is an example of a species that is very difficult to 
obtain an accurate set of measurements for, we can test a hypothesis about its formation
in wild type cells by looking at its predicted time course, Fig.~\ref{fig:predictions}. 
                                                                                
\begin{figure}[hbtp]
\vspace{2pt}
  
\centerline{\hbox{ \hspace{0.0in}
\hspace{0.3in}
\epsfxsize=3.0in
\epsffile{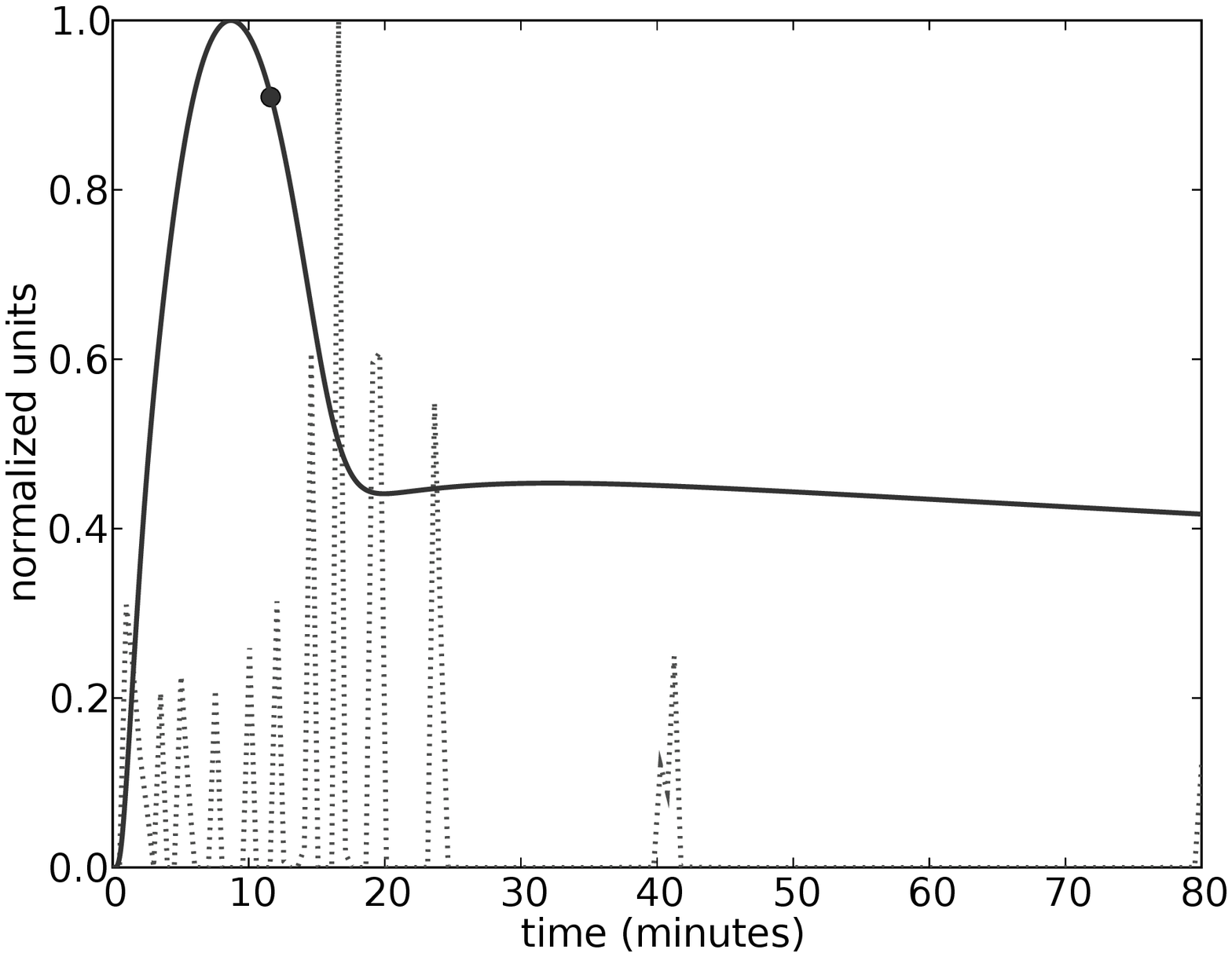}
\hspace{.2in}
\epsfxsize=3.0in
\epsffile{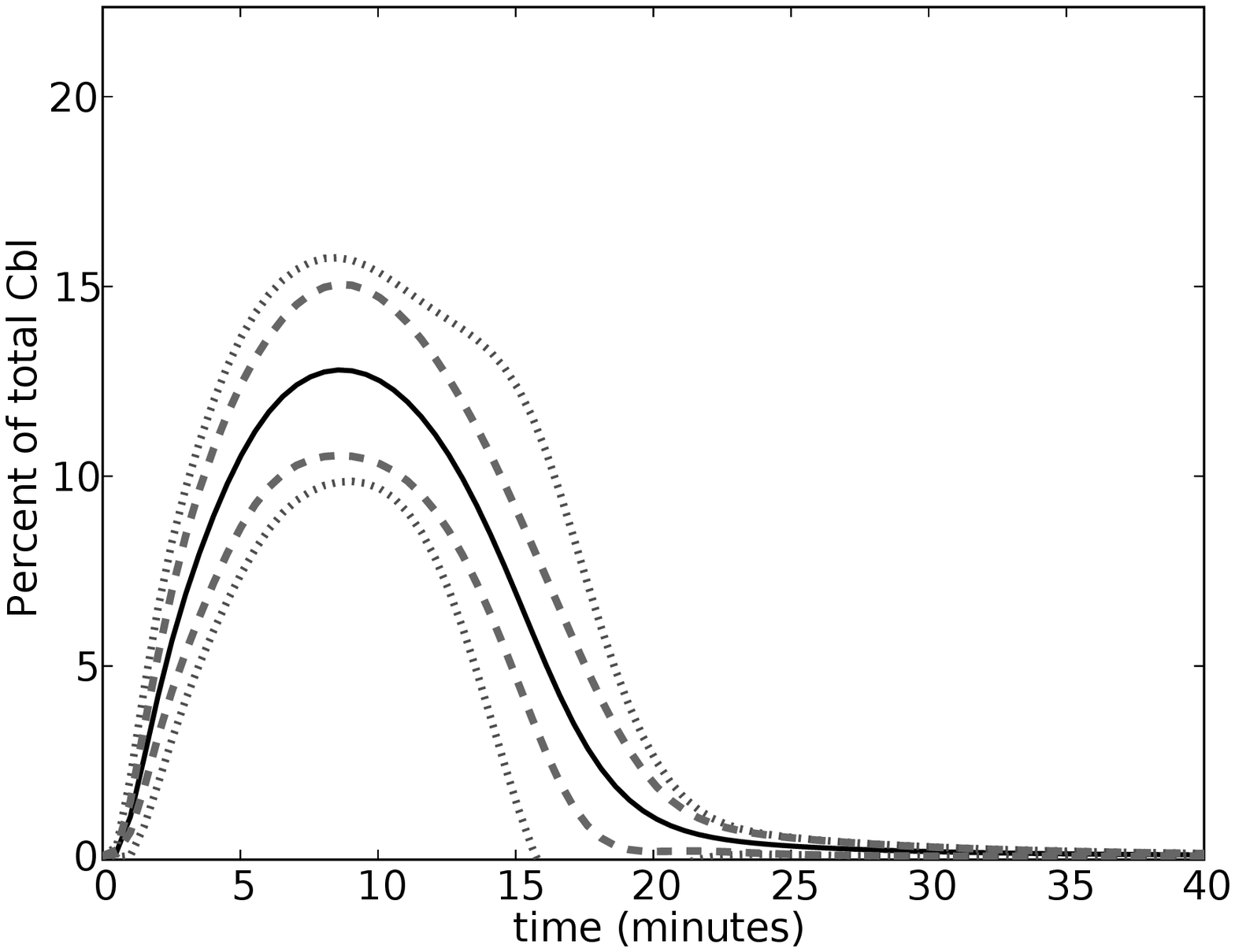}
					}
						  }
\vspace{5pt}
\hbox{\hspace{1.6in} (a) \hspace{3.1in} (b) \hspace{2.0in} }
					
\caption{(a) Trajectory of total active Cdc42 (solid line) with single 
		sequential design measurement (marked with a dot) and 
		approximate continuous design weights (dotted line)  
		to reduce the average variance of the prediction on the 
		Cool-1, Cbl, Cdc42 complex.
		The weights are optimized over 160 uniformly spaced hypothetical 
		measurements placed between 0 and 80 minutes on Cdc42. 
	   	(b) Shows the reduction in the original uncertainty bounds 
		resulting from the single measurement (dotted line) and the approximate
		continuous design measurement (dashed line) in (a). Compare with 
		Fig.~\ref{fig:predictions} before the addition of new measurements.
		}
\label{fig:designpoints}
\end{figure}

                                                                                
The relative amount of the triple complex is shown in Fig.~\ref{fig:predictions},
where the number of molecules of the triple complex has been scaled relative to 
the total level of Cbl.
\emph{Relative} levels of complexes and the times of formation/dissociation are more 
meaningful quantities than absolute numbers of molecules, which are merely rough
estimates used to initialize the simulations. 
The best fit trajectory for the triple complex suggests that at a 
maximum over 12\% of Cbl is sequestered in the complex which represents a 
significant proportion.
However the uncertainty bounds are too large to make this assertion; at the 
level of the lower bound, less than 4\% of Cbl is sequestered at a maximum,
and the triple complex dissociates within 15 minutes. This motivates the 
need for an optimal design approach. We define a criterion which is the 
average uncertainty in the prediction on the triple complex. We then
optimize this quantity using a sequential design
approach (therefore we need to perform only line minimizations in the 
time coordinate for each of the 11 measurable species in the system) and
follow up by finding an approximate optimal continuous design on that species. 
The results of such an analysis are shown in Fig.~\ref{fig:designpoints}. 

\begin{figure}[hbtp]
\begin{center}
\epsfxsize=6.0in
\epsfbox{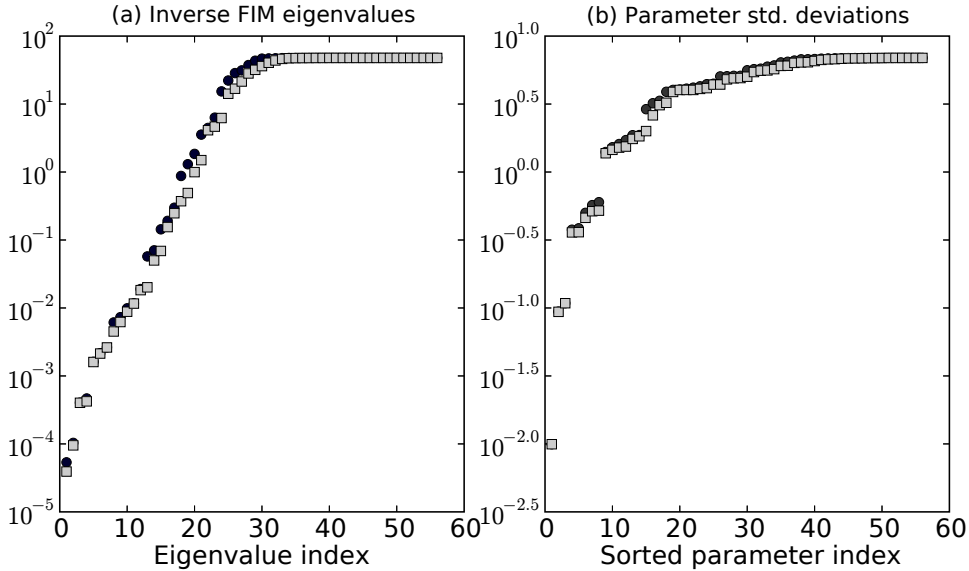}
\caption{(a) Eigenvalues of the approximate parameter covariance
matrix, $M^{-1}$, with (light squares) and without (dark circles) the optimally 
designed data to reduce the uncertainty of the 
triple complex trajectory. (b) Individual parameter standard
deviations, sorted from smallest to biggest with (light squares) and without (dark circles) the 
optimally designed data. Note that the cutoff in the spectrum of eigenvalues is due
to the prior information assumed on parameters ranges. Even with prior information, 40 of 
the 60 parameters have uncertainties corresponding to a greater than 
20-fold increase or decrease in their non-logarithmic values.}

\label{fig:eigenvalues}
\end{center}
\end{figure}

The most striking features of the optimal design results are that \\
\begin{enumerate}
\item a \emph{single} measurement on total active Cdc42 can significantly reduce the variance
we see in the prediction on the triple complex, as in Fig.~\ref{fig:designpoints} (b)
\item even though the approximate continuous design allows for
160 hypothetical measurements on the activity of Cdc42, the optimal design weights are
concentrated to just a dozen early time points. That is, by just taking a few 
measurements we can get a design very close to the optimal continuous design 
for measuring total active Cdc42.
\end{enumerate}
It is worth noting here that these extra measurements have little effect on the
parameter uncertainty. In Fig.~\ref{fig:eigenvalues} on the left, we show the eigenvalues of 
the approximate covariance matrix $M^{-1}$ both before and after the addition 
of the new data points. On the right is the square root of the diagonal elements of $M^{-1}$, 
giving the standard deviation in each parameter. As can be seen, the large 
parameter uncertainties are changed little after the addition of the optimal data points. 
In a sense, the underlying parameter uncertainty defined
by $M^{-1}$ in Eqn. \ref{predvariance}, although large in some directions, is 
mostly aligned with directions where the model sensitivity is small. Conversely, 
if we include hypothetical measurements on the binding and unbinding constants 
involved in forming the triple complex, we find only a negligibly small 
decrease in the uncertainty in the prediction of the triple complex (see supplementary
information). This is not so surprising when we understand that the uncertainty arises from 
the uncertainties in components of the system upstream of the triple complex; 
using parameter measurements alone, almost every rate constant in the system
would have to be measured accurately to constrain the prediction~\cite{Ryan}.
\subsubsection{New measurements on total active Cdc42}
\begin{figure}[hbtp]
  \vspace{2pt}
                                                                                
\centerline{\hbox{ \hspace{0.0in}
    \epsfxsize=3.0in
    \epsffile{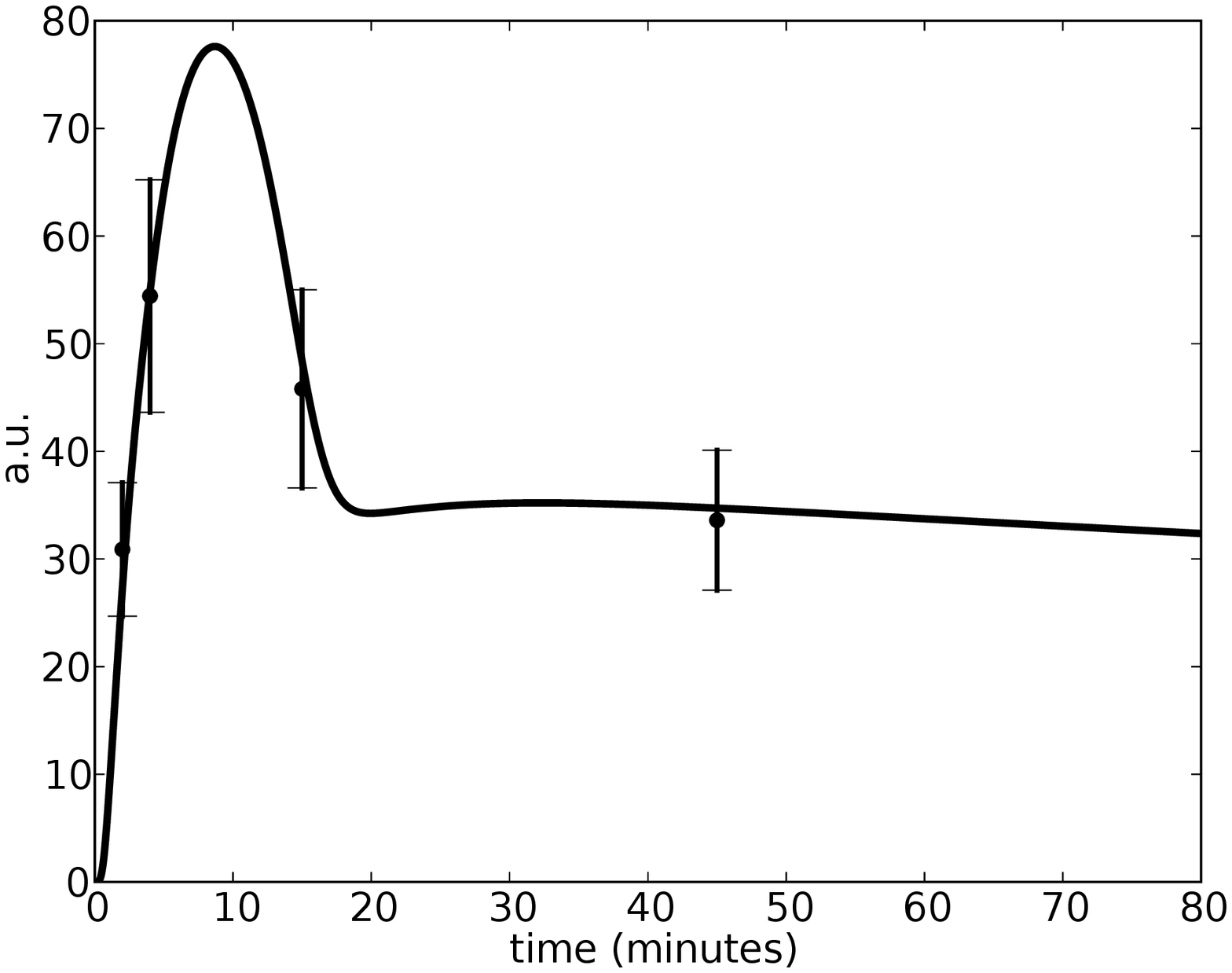}
    \hspace{0.3in}
    \epsfxsize=3.0in
    \epsffile{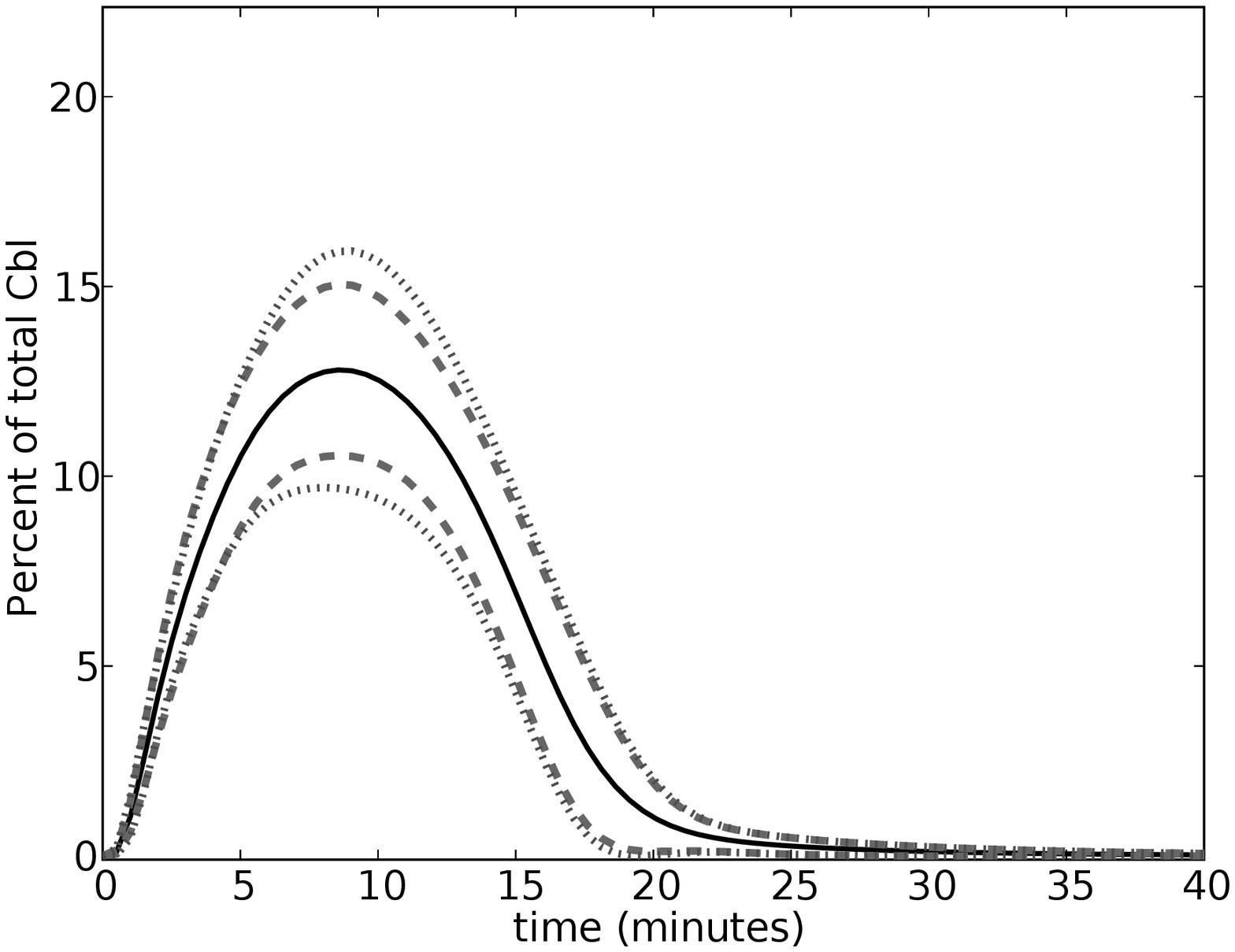}
   }
 }
\vspace{5pt}
  \hbox{\hspace{1.4in} (a) \hspace{3.1in} (b)}
                                                                                
    \caption{(a) Without refitting to the new total active Cdc42 data,
	our prediction matches the data using only a single
	multiplicative factor. {a.u.} = arbitrary units.
	(b) Reduced uncertainty on the time course of the active Cool, Cbl, and
	active Cdc42 complex for the optimal set of design points (dashed line) (same as 
	Fig.~\ref{fig:designpoints} (b) ) and for the real data (dotted line).
    }
  \label{fig:total42fit}
\end{figure}

Further measurements were made on total activated Cdc42 in the lab by Western 
blotting and with no refitting, our model was able to match the new data using 
a scale factor alone, see Fig.~\ref{fig:total42fit}. (However, we cannot consider
this as a validation of our model, since prior to the inclusion of the 
new data, the uncertainties on total activated Cdc42 were very large. Any 
experimental observations within the uncertainty 
bounds would be consistent with the model.) The 
uncertainties of the triple complex time course, given the real
data and the optimally weighted data, is shown in Fig.~\ref{fig:total42fit} (b).
Importantly, given that the measured activities of total Cdc42 were consistent
with the trajectory for the optimized set of parameters, the reduction in uncertainty
of the triple complex for the real data is comparable to that for the optimally selected data 
and we can make a firm conclusion that the triple complex does 
sequester significant amounts of the Cbl protein even in wild type cells after
EGF stimulation. Therefore it appears that the complex plays a part under  
normal conditions in the EGFR homeostasis. 
(Note that if the new data collected showed a very different
time course than in Fig.~\ref{fig:total42fit}, an additional re-optimization step 
would need to be performed before we could assess the prediction and uncertainties
for the triple complex.) 

\section{Discussion}
We have demonstrated that by quantitatively modeling the dynamics of EGFR
signaling and down-regulation in a mammalian cell line, we are led to 
incorporate interactions and modify existing reactions in order to reproduce the
experimental observations. Note that these interactions are not directly 
tested by experiments, but we can infer them from the existing data. 
This refinement of an existing model of interactions and parameters 
is one important aspect of the modeling effort and gives insight into the 
underlying dynamics. Of course, we recognize that the model as it stands will
only explain the behaviors observed in the data sets we have chosen. The
addition of new experiments that test for receptor signaling from
early endosomes~\cite{endosomalSig}, alternative endocytic mechanisms~\cite{difiore}, 
autocrine signaling~\cite{autocrine1,autocrine2} or the interactions between members of 
the erb-B family~\cite{erbBfamily}, for example, will require appropriate extensions 
of the mathematical model. 

The second part of the process is to make predictions on the unmeasured or 
unmeasurable species of the system, assuming that the model has been 
suitably refined. We suggest that for testable predictions to be made, 
uncertainty estimates need to be attached to them~\cite{kevin1}. 
In some cases the prediction uncertainties are rather small, despite large parameter 
uncertainty. On the other hand, if some predictions show large uncertainty, 
and involve species that are not directly measurable, we may then define a suitable 
design criterion and suggest new experimental measurements that need to be 
taken to reduce that uncertainty.
The results of such an analysis are promising, in that we find a rather 
small number of measurements (realistic to perform with standard molecular 
biology techniques) need be taken to begin to make predictions with 
good precision. Given such measurements on the EGFR system, we see that 
the triple complex of active Cool-1, Cbl and active Cdc42 does indeed 
form in appreciable quantities in wild type cells and we also get an estimate 
for the time of formation and dissociation.

More generally, we believe that experimental design for reducing prediction 
uncertainties can play an important role in the iterative process of model 
refinement and validation and can be used in the testing of biological hypotheses.

\bibliographystyle{siam}
\end{document}


\section{Supplementary information}
\subsection{Experimental data}
The experimental data are drawn from literature sources~\cite{3T3expt1,3T3expt2,3T3expt3} 
and collected in the lab of the co-authors (D.B., Q.F. and R.A.C.)~\cite{qiyulatest}. 
All measurements were performed  
using Western blots assays on NIH-3T3 cells and were quantified using
densitometry. Within the simulation 
all species were represented in numbers of molecules per cell. As many 
of the reactions are restricted to the cell membrane, the numbers quoted
can be thought of as effective amounts associated with the membrane. 
As we are only interested in relative amounts and relative changes of protein
level or activity, the absolute numbers used are of less importance and 
should not be interpreted literally. Approximate 
receptor numbers were either reported with the source of the experiment or 
were left as free parameters in the model.  
EGF molecule number per cell was estimated assuming one million cells per
dish and an aliquot of EGF solution of volume 3nL. 

All experimental conditions involved serum starvation followed by EGF stimulation 
of various levels. The Western blots were quantified by measuring average 
pixel density in each lane. Error bars (when not available through replicates) were
assigned typically as 20\% of nominal value. 

The full set of numbered experiments, corresponding to fits shown in 
Fig.~\ref{fig:fits}, are as follows :
\begin{enumerate}
\item Percentage of receptors remaining on cell surface 
(bound or free of ligand) as a function of incubation time with 100nM EGF. 
100,000 receptors (transfected) reported initially~\cite{3T3expt1}.
\item Percentage of receptors remaining on cell surface as a function of 
incubation time with 100ng/ml EGF. 100,000 receptors (transfected) reported
initially~\cite{3T3expt2}.
\item Percentage of receptors remaining on cell surface as a function of 
incubation time with 100ng/ml EGF. 275,000 receptors (transfected) reported
initially~\cite{3T3expt2}.
\item Percentage of surface, internal and degraded EGF after pre-loading
the receptors with EGF in conditions that prevent internalization and 
then allowing internalization at time zero. There is no EGF exposure 
apart from preloaded amounts. 100,000 receptors (transfected) reported
initially~\cite{3T3expt1}.
\item Total level of EGFR as a function of incubation time with 100ng/ml EGF.
3 experimental conditions: endogenous levels of Cdc42, transfection of the 
Cdc42F28L fast cycler, transfection of the Cdc42($\Delta$L8/F28L) fast cycler 
which is Cool-1 binding defective. 8000 receptors (endogenous) reported 
initially~\cite{3T3expt3}.
\item Phosphorylation of Cool-1 (endogenous levels) after incubation with
100ng/ml EGF for the given times. Number of EGFR receptors (transfected)
is a free parameter~\cite{qiyulatest}.
\item Phosphorylation of Cool-1 (transfected) after incubation with 
100ng/ml EGF for the given times. Number of Cool-1 molecules taken as 
10 fold over normal. Number of EGFR receptors (transfected) is a 
free parameter~\cite{qiyulatest}.
\item Phosphorylation of Erk after incubation with 100ng/ml EGF for the 
given times. 2 experimental conditions: endogenous levels of Cdc42 or 
an siRNA knockdown of Cdc42. The knockdown was estimated to reduce the
Cdc42 levels to 18 percent of normal~\cite{qiyulatest}. EGFR was at endogenous levels 
(assumed to be 8000 receptors as in \cite{3T3expt3}).
\item Phosphorylation of Erk after incubation with 100ng/ml EGF for the
given times. 2 experimental conditions: endogenous levels of Cool-1 or 
an siRNA knockdown of Cool-1. The knockdown was estimated to reduce the
Cool-1 levels to 26 percent~\cite{qiyulatest}. EGFR was at endogenous levels
(assumed to be 8000 receptors as in \cite{3T3expt3}).
\item Phosphorylation of Cool-1 after incubation with 100ng/ml EGF for 
a range of times. EGFR number (transfected) is a free parameter. This
experiment differs from the previous endogenous Cool-1 activation assay 
in that early time points are measured~\cite{qiyulatest}. 
\item Activation of Cdc42 (total level of Cdc42 GTP-bound) after 
incubation with 100ng/ml EGF~\cite{qiyulatest}. EGFR was at endogenous levels 
(assumed to be 8000 receptors as in \cite{3T3expt3}).
\end{enumerate}

\subsection{Model details}
\begin{figure}[hbtp]
\begin{center}
\epsfxsize=6.0in
\epsfbox{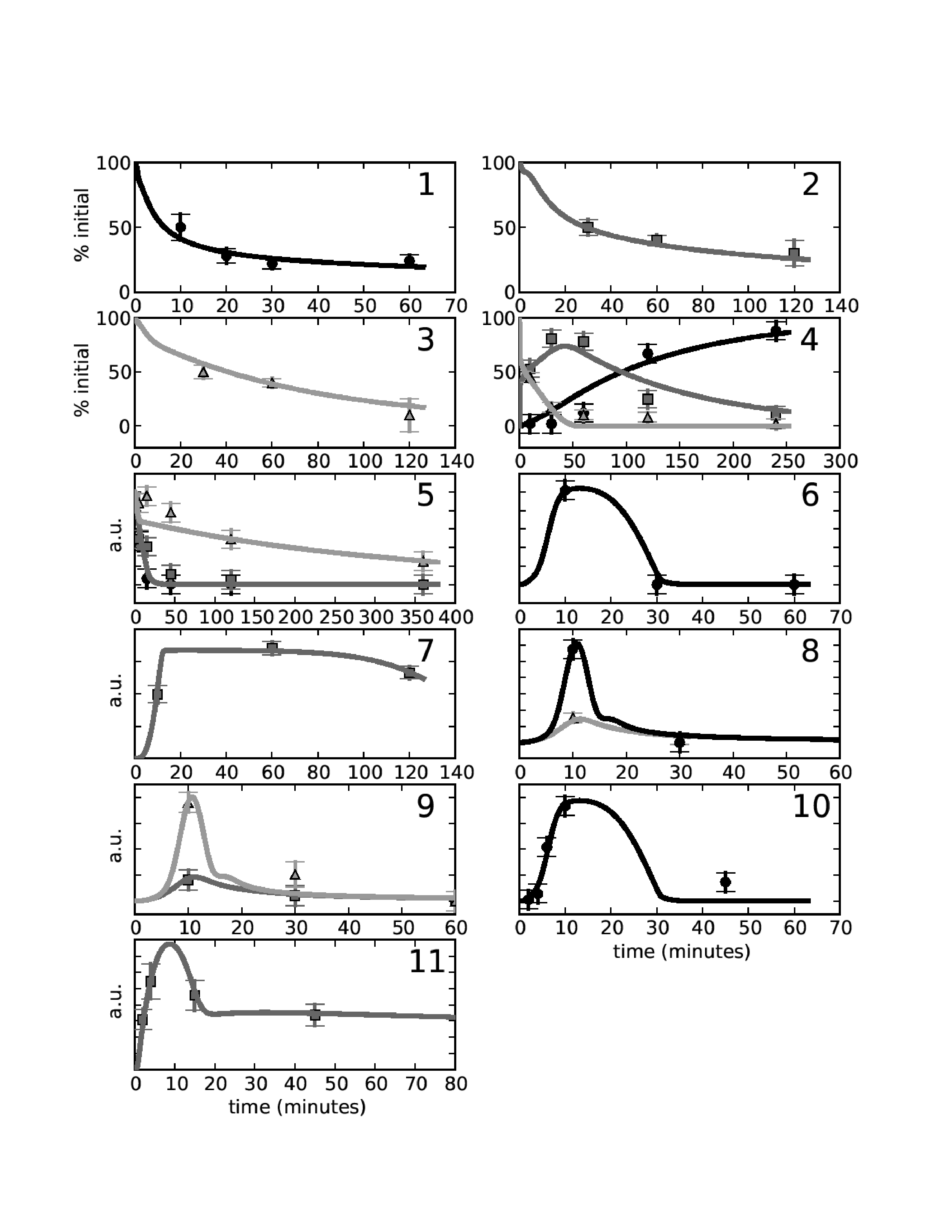}
\caption{Experimental data and fits for all the experiments described in the 
text. Experiment 4: surface EGF (light triangles), internal EGF (gray squares), 
degraded EGF (dark circles). Experiment 5: Cdc42(F28L) overexpression (light 
triangles), Cdc42(F28L/DL8) (gray squares), control (black circles). Experiment 8: 
Cdc42 siRNA (light triangles), control (black circles). Experiment 9: Cool-1 siRNA
(gray squares), control (light triangles). All other experiments measure a single
species for a single experimental condition described in text.}
\label{fig:fits}
\end{center}
\end{figure}

The model as shown in Fig. 1 of the main text involves 41 distinct dynamical 
variables and 53 unknown rate and Michaelis-Menten constants. 
Additionally, it was found useful to include the total number of Cool, 
Cbl and Cdc42 molecules as fitted variables. Thus the optimization problem of 
minimizing the least squares cost function involves 56 free parameters. 
Optimization was primarily carried out using a Levenberg-Marquardt 
method~\cite{trustregion}, working in 
the natural logarithm of the biochemical constants to enforce the positivity constraint. The 
fits to the experimental data are shown in Fig.~\ref{fig:fits}, including the last 
experiment performed on active Cdc42 which needed no re-optimization.

The system was modeled using SloppyCell, 
available at {\tt{http://sloppycell.sourceforge.net}}.
SloppyCell is a Python-based modeling environment which facilitates the development, simulation
and optimization of biochemical networks. The Systems Biology Markup Language (SBML) 
format file~\cite{Hucka2003} for the model 
(interpretable by SloppyCell and many other biochemical network 
simulators) and best fit parameter sets can be obtained from the first author.

\subsection{Parameter sensitivities}
\begin{figure}[hbtp]
\begin{center}
\epsfxsize=4.0in
\epsfbox{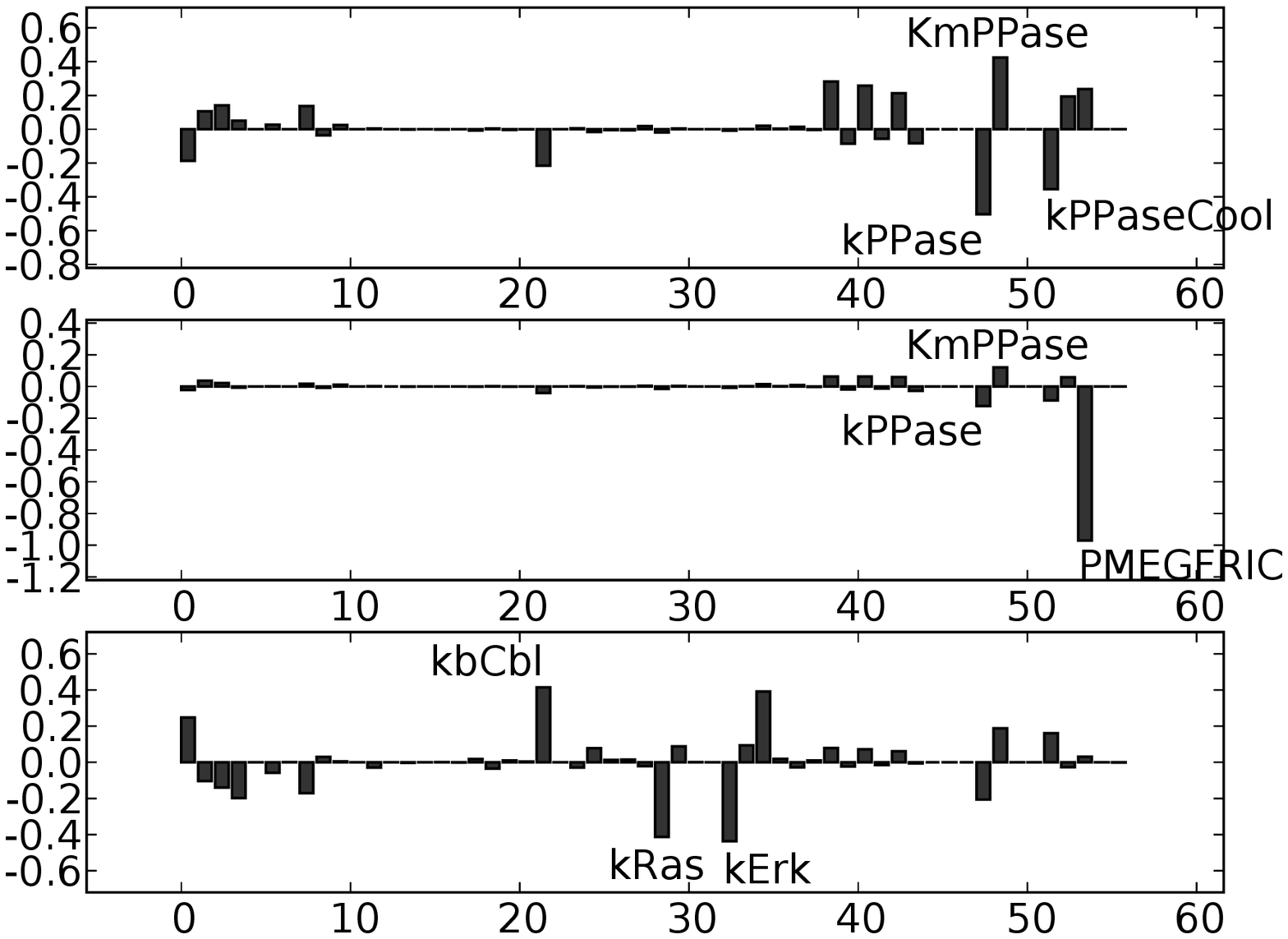}
\caption{Components of the first three eigenvectors of the FIM evaluated at the
best fit for the EGFR model}
\label{fig:eigenvectors}
\end{center}
\end{figure}

\begin{figure}[hbtp]
\begin{center}
\epsfxsize=4.0in
\epsfbox{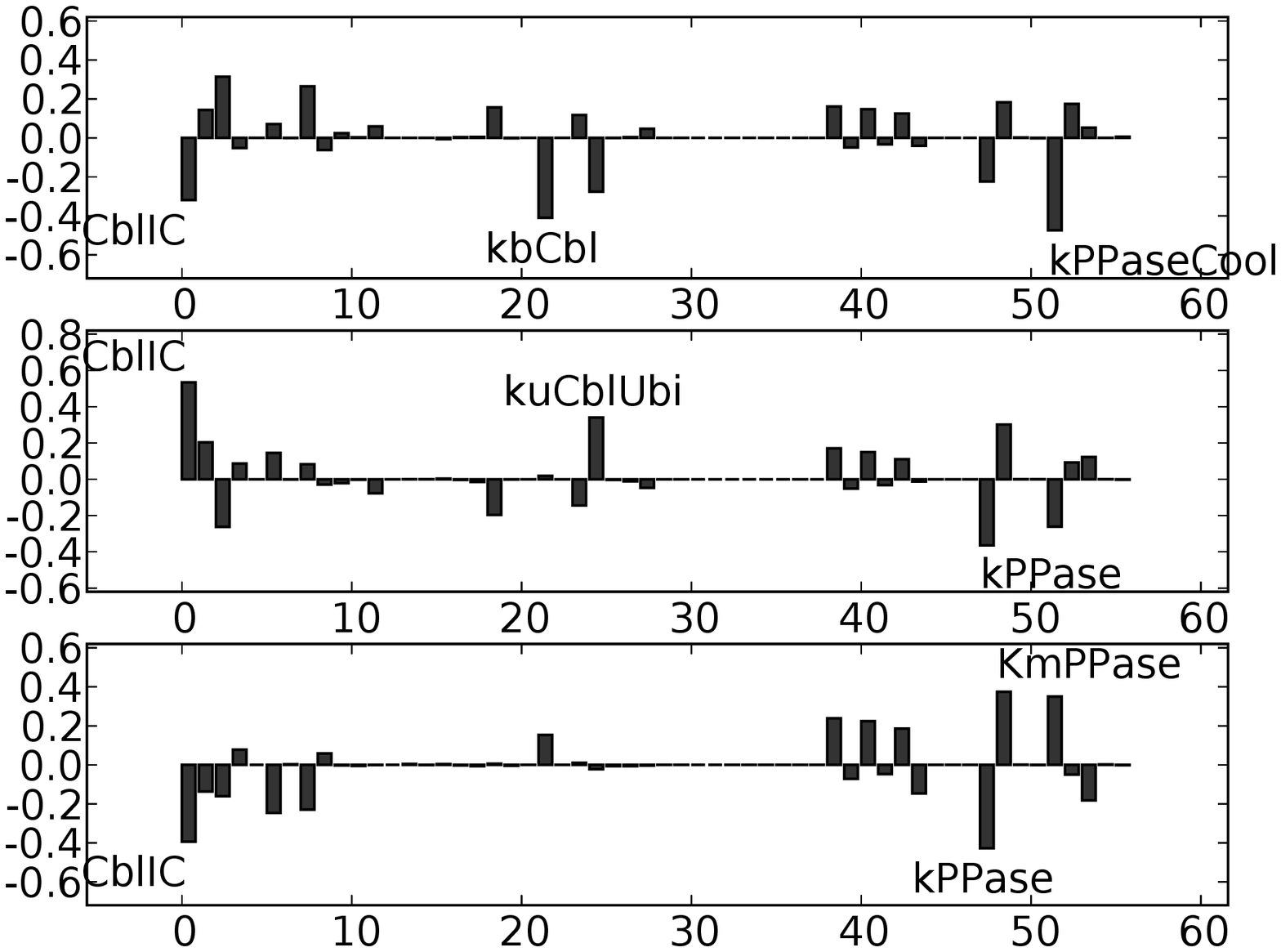}
\caption{Components of the first three eigenvectors of the FIM evaluated at the
best fit for the EGFR model with simulated data on all species at 2 minute
intervals}
\label{fig:eigenvectorsPerfect}
\end{center}
\end{figure}

A common technique in model exploration is sensitivity analysis --- determining which 
directions in parameter space are dominant in controlling the system behavior (as 
measured by the fit to the data in this case). Fig.~\ref{fig:eigenvectors} shows the 
first three eigenvectors of the FIM which show the directions in parameter space that are 
most constrained by the experimental data.
Note that the most sensitive components of the pathway appear to be related
to the action of the phosphatases: a generic phosphatase for Cbl, FAK, Erk 
and the EGFR and a specific phosphatase for Cool-1 (SHP-2). Furthermore, 
the activation pathway of Cool-1 (through FAK and Src) appears to be 
quite constrained by the experimental data. Whether this is just a 
function of this particular set of experimental data, or a feature of the 
model behavior we can test by simulating data on \emph{all} species in the
system. Thus we are not biasing the sensitivity analysis towards species 
that have the most data available. The principal 3 eigenvectors for the
system with data simulated for all species at 2 minute timepoints is 
shown in Fig.~\ref{fig:eigenvectorsPerfect}. The largest
component in the first eigenvector is still the rate constant for the 
Cool phosphatase and the rate constants of the generic phosphatase are
large components in the second eigenvector (the first eigenvector
for the simulated data has a dot product of .76 with that for the 
real data).

\subsection{Posterior parameter distribution}
\begin{figure}[hbtp]
\begin{center}
\epsfxsize=4.0in
\epsfbox{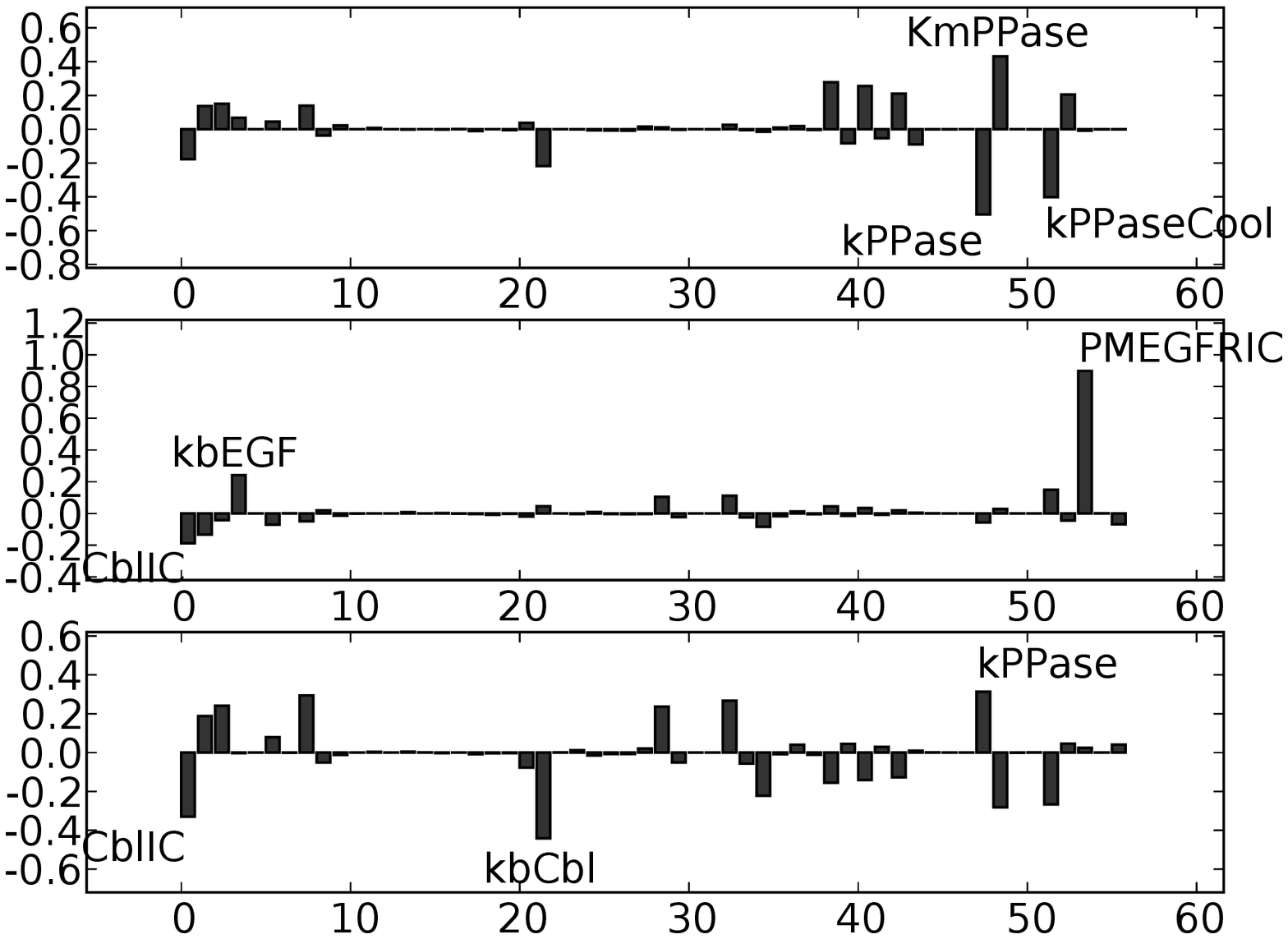}
\caption{The last three principal component axes, in reverse order, 
(most tightly constrained directions) for a sample of approximately 5000 parameter 
sets sampled from the posterior distribution.}
\label{fig:pcavectors}
\end{center}
\end{figure}

Another technique to measure the influence of parameter directions on 
the behavior of the system is to take a Bayesian approach and directly sample 
the likelihood function for the data~\cite{tarantola,kevin1}. By applying 
a Markov Chain Monte Carlo (MCMC) algorithm we can build up a sample of 
parameter sets from the posterior distribution. We have used the Metropolis
Hastings method and run eight MCMC chains for approximately 170 hours on 
the model with all the experimental data (including the last experiment
on total active Cdc42).
From the resulting parameter set sample, we removed correlated filtered 
out correlated members, leaving approximately 5000 parameter sets as a 
independent draws from the posterior distribution.
The posterior sample will be more constrained in some directions than 
in others. The most tightly constrained directions (as determined by 
doing a Principal Component Analysis (PCA) on the cloud of parameter sets) 
are directions that most affect the behavior of the model. The 3 PCA axes 
corresponding to the most constrained directions are shown in 
Fig.~\ref{fig:pcavectors}.
The last principal component axis has a dot product of .96 with the first
eigenvector of the FIM, demonstrating that the linear approximation underlying
the FIM is quite accurate, at least in determining the most constrained 
direction.

\begin{figure}[hbtp]
\begin{center}
\epsfxsize=6.0in
\epsfbox{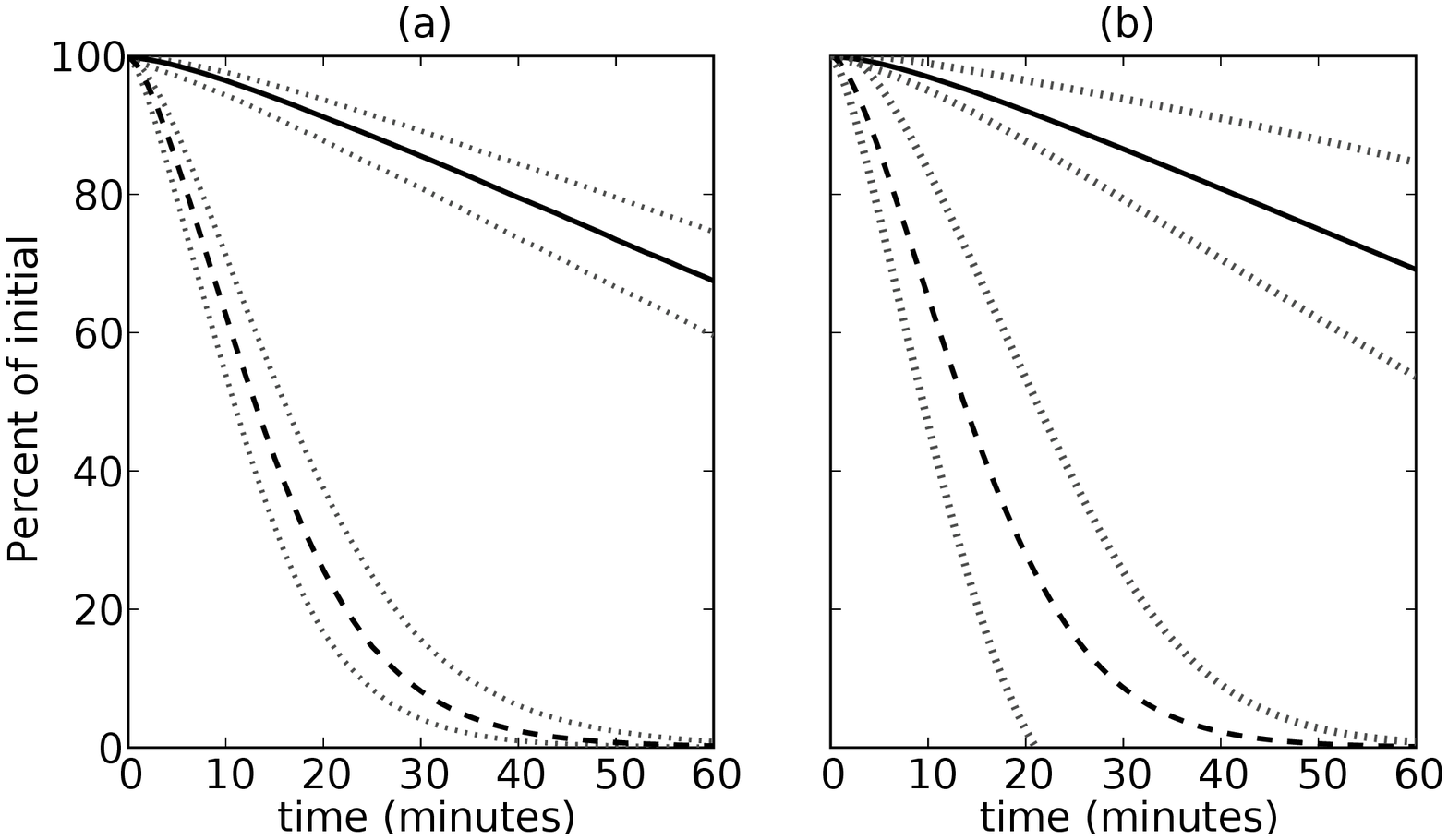}
\caption{The total surface receptor numbers after EGF stimulation in stably 
transfected v-Src cells. 
The two experimental conditions are endogenous levels of Cool-1 (dashed curve) or 
overexpressed Cool-1 (solid curve). (a) The total surface receptor trajectory is 
a median of all the trajectories in the posterior distribution. 
The uncertainties are the boundaries of the 68\% credible regions (dotted lines) to
correspond with the one standard deviation errors shown in the main text. (b) The 
linearized estimates, shown for comparison.}
\label{fig:predensembles}
\end{center}
\end{figure}

\begin{figure}[hbtp]
\begin{center}
\epsfxsize=6.0in
\epsfbox{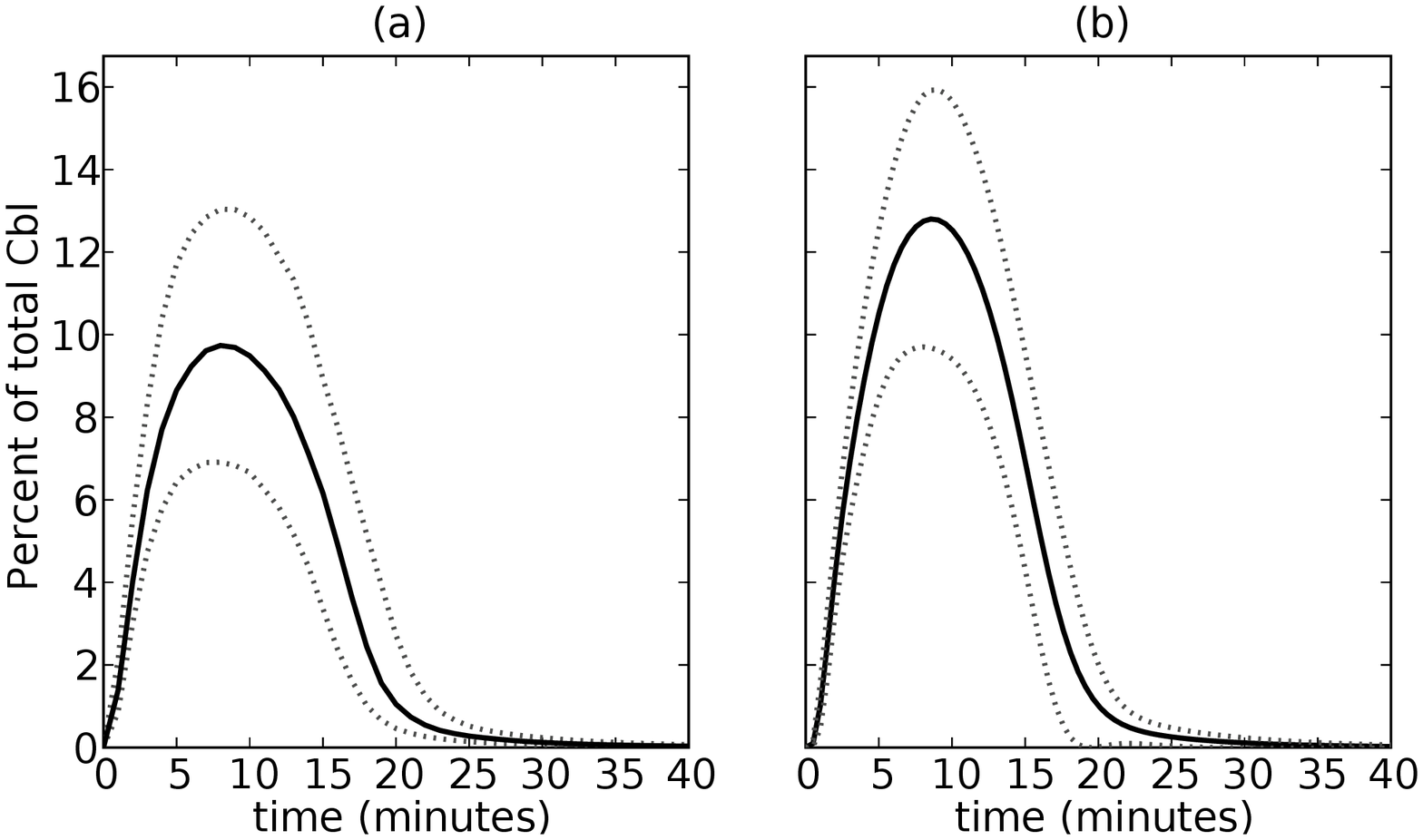}
\caption{The level of the triple complex in wild type cells with endogenous 
EGFR. 
(a) Median trajectory (solid curve) and the uncertainties given by the 
boundaries of the 68\% credible region (dotted lines). (b) The linearized 
estimates, shown for comparison. Note that the best fit trajectory in (b) is
noticeably different from the median trajectory shown in (a)}
\label{fig:tripcompens}
\end{center}
\end{figure}

The Bayesian approach also provides us with a tool for generating a 
distribution of predictions from the posterior distribution 
of parameter sets. As these predictions are derived from a sampling of the 
full likelihood function (compared to the Gaussian approximation assumed in the
Fisher Information Matrix approach) we would expect it to give a more accurate 
estimate of the true uncertainties. For comparison we show in 
Fig.~\ref{fig:predensembles} and Fig.~\ref{fig:tripcompens} the predictions on total
surface EGFR in v-Src cells and triple complex levels in wild type cells, using the
Bayesian error estimates.  

The uncertainties are sufficiently similar between the two approaches that
we can justify the methods used in the main text and can confirm the conclusions
made there.
\subsection{Parameter information and prediction uncertainty}
\begin{figure}[hbtp]
\begin{center}
\epsfxsize=4.0in
\epsfbox{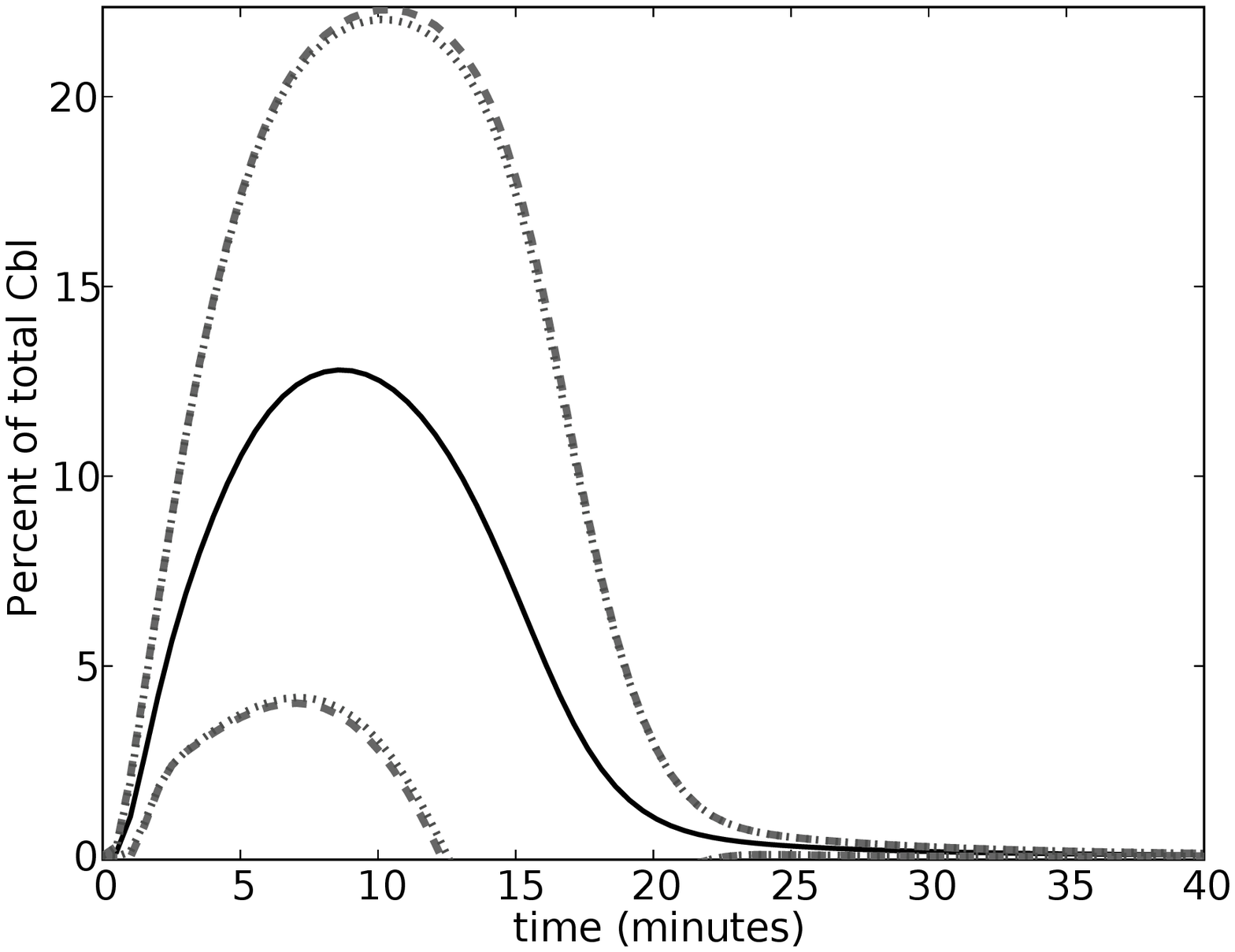}
\caption{The uncertainties on the prediction of the triple complex with no 
formation/dissociation rate constant information (dashed line) and 
assuming the relevant rate constants are known to 
within 5\% (dotted line).}
\label{fig:kukbnogood}
\end{center}
\end{figure}

As mentioned in the main text, the addition of the new data had little effect on the 
parameter uncertainty in the system but significantly reduces the uncertainty in the 
prediction of the time course for the triple complex. Conversely, assuming precise  
values for the rate constants involved in the formation and dissociation of the triple
complex (binding and unbinding rate constants) does little to change the uncertainty
in the prediction. This is shown in Fig.~\ref{fig:kukbnogood}.
The reason for the apparent insignificance of the rate constant information is that
the underlying uncertainty in the trajectory of the triple complex derives from the
uncertainty in the \emph{components} from which it is comprised, rather than from the 
uncertainty in the rate constants for formation and dissociation.

\bibliographystyle{siam}